\documentclass[wes, manuscript]{copernicus}
\usepackage{booktabs}
\usepackage{footnote}

\begin{document}
\nolinenumbers

\title{Operational-Based Annual Energy Production Uncertainty: Are its Components Actually Uncorrelated?}

\Author[1]{Nicola Bodini}{}
\Author[1]{Mike Optis}{}

\affil[1]{National Renewable Energy Laboratory, Golden, Colorado, USA}

\correspondence{Mike Optis (mike.optis@nrel.gov)}

\runningtitle{Correlation of uncertainties in wind plant operational annual energy production estimates}

\runningauthor{N. Bodini and M. Optis}

\received{}
\pubdiscuss{} 
\revised{}
\accepted{}
\published{}

\firstpage{1}

\maketitle

\begin{abstract}
Calculations of annual energy production (AEP) from a wind power plant---whether based on preconstruction or operational data---are critical for wind plant financial transactions. The uncertainty in the AEP calculation is especially important in quantifying risk and is a key factor in determining financing terms. A popular industry practice is to assume that different uncertainty components within an AEP calculation are uncorrelated and can therefore be combined as the sum of their squares. We assess the practical validity of this assumption for operational-based uncertainty, by performing operational AEP estimates for more than 470 wind plants in the United States, mostly in simple terrain. We apply a Monte Carlo approach to quantify uncertainty in five categories: revenue meter data, wind speed data, regression relationship, long-term correction, and future interannual variability. We identify correlations between categories by comparing the results across all 470 wind plants. We observe a positive correlation between interannual variability and the linearized long-term correction; a negative correlation between wind resource interannual variability and linear regression; and a positive correlation between reference wind speed uncertainty and linear regression. Then, we contrast total operational AEP uncertainty values calculated by omitting and considering the correlations between the uncertainty components. We quantify that ignoring these correlations leads to an underestimation of total AEP uncertainty of, on average, 0.1\%, and as large as 0.5\% for specific sites. Although these are not large increases, these would still impact wind plant financing rates; further, we expect these values to increase for wind plants in complex terrain. Based on these results, we conclude that correlations between the identified uncertainty components should be considered when computing the total AEP uncertainty.
\end{abstract}

\copyrightstatement{This work was authored by the National Renewable Energy Laboratory, operated by Alliance for Sustainable Energy, LLC, for the U.S. Department of Energy (DOE) under Contract No. DE-AC36-08GO28308. Funding provided by the U.S. Department of Energy Office of Energy Efficiency and Renewable Energy Wind Energy Technologies Office. The views expressed in the article do not necessarily represent the views of the DOE or the U.S. Government. The U.S. Government retains and the publisher, by accepting the article for publication, acknowledges that the U.S. Government retains a nonexclusive, paid-up, irrevocable, worldwide license to publish or reproduce the published form of this work, or allow others to do so, for U.S. Government purposes.}

\section{Introduction}

Calculations of wind plant annual energy production (AEP)---whether based on preconstruction data before a wind power plant is built or on operational data after a wind plant has started its operations---are vital for wind plant financial transactions. Preconstruction estimates of AEP are needed to secure and set the terms for new project financing, whereas operational estimates of long-term AEP are required for important wind plant transactions, such as refinancing, purchasing/selling, and mergers/acquisitions. The need for AEP analyses of wind plants is increasing because global wind capacity increased to 539~GW in 2017, representing 11\% and 91\% increases over 1-year and 5-year periods, respectively; capacity is expected to increase by another 56\%, to 841 GW, by 2022 \citep{gwec2017}. In the United States, wind plants generated more than 300,000 \unit{GWh} in 2019, about 7.5\% of the total U.S. electricity generation from utility-scale facilities that year, with a 50\% increase over a 6-year period \citep{us2020monthly}.

This rapid growth of the wind energy industry is putting an increased spotlight on the accuracy and consistency of AEP calculations. For preconstruction AEP estimates, there has been considerable movement toward standardization. The International Energy Commission (IEC) is currently developing a standard  \citep{iec15} and there have long been guidance and best practices available \citep{brower2012}. By contrast, long-term operational AEP estimates do not have such extensive guidance or standards. Only limited standards covering operational analyses exist; \cite{iec12} addresses turbine power curve testing, and \cite{iec26} addresses the derivation and categorization of availability loss metrics. However, to our knowledge, there are no standards and very limited published guidance on calculating long-term AEP from operational data. Rather, documentation seems to be limited to a consultant report \citep{lindvall2016}, an academic thesis \citep{khatab2017}, and limited conference proceedings \citep{cameron2012,lunacek2018}. 

Documentation and standards for preconstruction AEP methods are of limited use for operational-based AEP methods, given the many differences between the two approaches. In general, operational AEP calculations are simpler than preconstruction estimates because actual measurements of wind plant power production at the revenue meter replace the complicated preconstruction estimate process (e.g., meteorological measurements, wind and wake-flow modeling, turbine performance, estimates of wind plant losses). However, the two methods do share several similarities, including regression relationships between on-site measurements and a long-term wind speed reference, the associated long-term (windiness) correction applied to the on-site measurements, estimates of future interannual variability, and estimates of uncertainty in the resulting AEP calculation. The shared components between operational AEP calculations and preconstruction estimates \citep{iec15} are listed in Table \ref{tab:unc_cat}.

\begin{table*}[h]
\centering
\begin{tabular}{p{5cm}p{10cm}} 
\toprule
Uncertainty component & Description \\ 
\midrule
On-site measurements & Measurement error in met mast wind speeds (preconstruction) or power at the revenue meter (operational) \\
{Reference wind speed data} & Measurement or modeling error in measured or modeled long-term reference wind speed data \\
Losses & Error in estimated or reported availability and curtailment losses \\
Regression & Sensitivity in the regression relationship between on-site measurements and reference wind speeds \\
Long-term (windiness) correction & Sensitivity in the long-term correction applied to the regression relationship between on-site measurements and reference wind speeds \\
Interannual variability of resource & Sensitivity in future energy production because of resource variability \\
\bottomrule
\\
\end{tabular}
\caption{Main Sources of Uncertainty in an AEP Estimate.}
\label{tab:unc_cat}
\end{table*}

The uncertainty values from each component listed in Table \ref{tab:unc_cat} must be combined to produce a total estimate of AEP uncertainty. While general guidelines on how to combine (measurement) uncertainty components exists \citep{iso1995guide} and can be applied to this task, we found no specific guidance in the literature for combining uncertainty components in an operational AEP estimate. On the other hand, considerable guidance exists for combining preconstruction AEP uncertainties \citep{lackner2007, brower2012, vaisala2014, kalkan2015, clifton2016}. In every case, recommended best practices assume that all uncertainties, $\sigma_\text{i}$, are uncorrelated and can therefore be combined using a sum of squares approach to give the total AEP uncertainty, $ \sigma_{\text{tot,uncorr}}$:
\begin{equation}
    \sigma_{\text{tot,uncorr}} = \sqrt{\sum_\text{i=1}^\text{N} \sigma_\text{i}^2}
    \label{unc_unc}
\end{equation}

To better understand how uncertainties are combined in long-term operational AEP calculations, we reached out to several wind energy consultants who regularly perform these analyses. These conversations revealed that uncertainties in a long-term operational AEP calculation are also assumed uncorrelated and combined using Equation \ref{unc_unc}.

\subsection{Goal of Study}

The purpose of this study is to examine the extent to which the assumption of uncorrelated uncertainties---and, therefore, the combination of those uncertainties through a sum of squares approach---is accurate and appropriate for operational AEP calculations. Specifically, this study aims to identify potential correlations between AEP uncertainty components, using data for over 470 wind plants. While in the analysis we focus on operational AEP calculation, we expect that the results from this analysis---namely, the potential identification of correlated uncertainty components---can be equally relevant for informing and improving preconstruction AEP methods.

In Section 2, we first describe the data sources used in this analysis (wind plant operational data and reanalysis products), the Monte Carlo approach to quantify single uncertainty components in operational AEP, and the approaches used to combine these uncertainty components. Section 3 presents the main results of our analysis in terms of uncertainty contributions and correlation among the different components. We conclude and suggest future work in Section 4.

\section{Data and Methods}

\subsection{Wind plant Operational Data and Reanalysis Products}

Operational wind plant energy production data for this analysis are obtained from the publicly available Energy Information Administration (EIA) 923 database \citep{eia18}. This database provides reporting of monthly net energy production from all power plants in the United States, including wind plants. More than 670 unique wind plants are available from this data set. 

Long-term wind speed data (needed to perform the long-term or windiness correction in an AEP estimate) are used from three reanalysis products over the period of January 1997 through December 2017:
\begin{itemize}
    \item Version 2 of the Modern-Era Retrospective analysis for Research and Applications (MERRA-2, \citet{gelaro2017modern}). We specifically use the M2T1NXSLV data product, which provides diagnostic wind speed at 50 m above ground level (AGL), interpolated from the lowest model level output (on average about 32 m AGL), using Monin Obukhov similarity theory. Data are provided at an hourly time resolution. 
    
    \item The European Reanalysis Interim (ERA-Interim) data set \citep{dee2011era}. We specifically use output at the 58$^\text{th}$ model level, which on average corresponds to a height of about 72 m AGL. Data are provided at a 6-hourly time resolution.
    
    \item The National Centers for Environmental Prediction v2 (NCEP-2) data set \citep{saha2014ncep}. We specifically use diagnostic wind speed data at 10 m AGL. Data are provided at a 6-hourly time resolution.
\end{itemize}

The wind speed data are density-corrected at their native time resolutions to correlate more strongly with wind plant power production (i.e., higher-density air in winter produces more power than lower-density air in summer, wind speed being the same):
\begin{equation}
    U_{\text{dens,corr}} = U \left(\frac{\rho}{\rho_{\text{mean}}} \right) ^{1/3}
    \label{den_corr}
\end{equation}
where $U_{\text{dens,corr}}$ is the density-corrected wind speed, $U$ is the wind speed, $\rho$ is air density (calculated at the same height as wind speed), $\rho_{\text{mean}}$ is the mean density over the entire period of record of the reanalysis product, and the exponent $1/3$ is derived from the basic relationship between wind power and wind speed cubed \citep{manwell2010wind}. To calculate air density at the same height as wind speed, we first extrapolate the reported surface pressure to the wind speed measurement height, assuming hydrostatic equilibrium \citep{atmosphere1975iso}:
\begin{equation}
    p = p_{\text{surf}} \: \exp \left[ \frac{g z}{RT_{\text{avg}}} \right]    
\end{equation}
where $p$ is the pressure at the wind speed measurement height, $p_{\text{surf}}$ is the surface pressure, $g$ is the acceleration caused by gravity, $z$ is the wind speed measurement height, $R$ is the gas constant, and $T_{\text{avg}}$ is the average temperature between the reported value at 2 m AGL and at the wind speed measurement height.

\begin{figure*}[t]
	\centering
	\includegraphics[width=16cm]{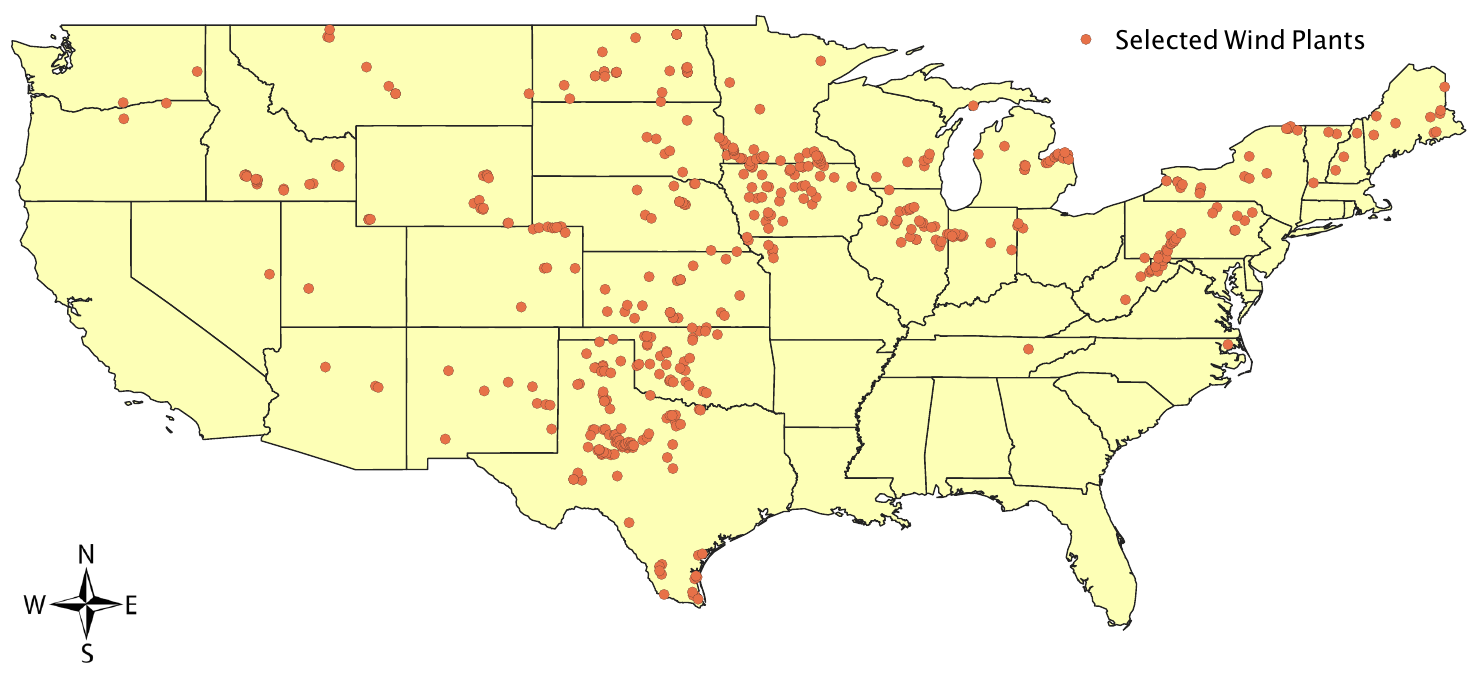}
	\caption{Map of the 472 wind plants that were considered in this study.}
	\label{Map_Sel}
\end{figure*}

To lessen the impact of limited and/or poor-quality data on the results of our analysis, we filter for wind plants with a moderate-to-strong correlation with all three reanalysis products ($R^2 >$ 0.6). About 25\% of the EIA wind plants are discarded with this filter. We also impose a threshold of eight months of wind plant data availability in order to investigate uncertainty as it relates to a low number of data points---but not so low as to make the use of a regression relationship questionable. A total of 472 wind plants are kept for the final analysis, and their locations are shown in Figure \ref{Map_Sel}. Because obtaining an accurate representation of wind data in complex terrain by reanalysis products is challenging \citep{shravan2009comparision}, most of the selected wind plants are located in the Midwest and Southern Plains. Notably, no wind plants in California pass the filtering criteria because they are predominately located in areas with thermally driven wind regimes, such as Tehachapi Pass, where coarse-resolution reanalysis products are poor predictors of wind energy production.

\begin{figure*}[t]
	\centering
	\includegraphics[width=12cm]{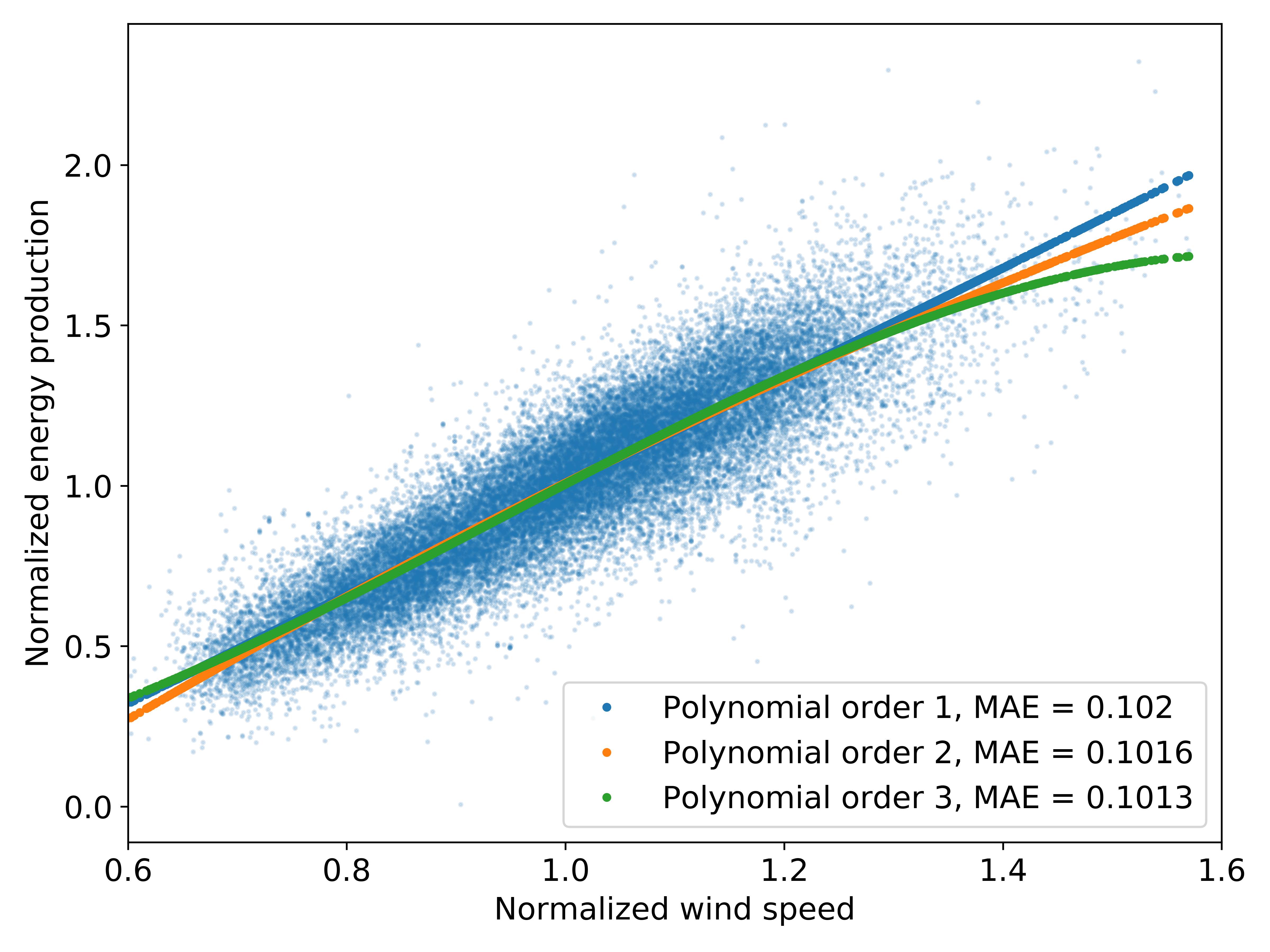}
	\caption{Scatterplot between normalized MERRA-2 density-corrected monthly wind speed and monthly energy production across all 472 selected sites, and linear, quadratic, and cubic best-fit lines.}
	\label{linear_reg}
\end{figure*}

The fundamental step in an AEP calculation involves a regression between density-corrected wind speed (here, from the reanalysis products) and energy production (here, from the EIA 923 database). To investigate whether a simple linear function can be assumed to express the relationship between density-corrected wind speed and wind plant energy production when considering monthly data, we show a scatterplot between MERRA-2 density-corrected monthly wind speed and monthly energy production across all 472 sites in Figure \ref{linear_reg}. For each site, data have been normalized by the respective site mean. We show best-fits using a linear, quadratic, and cubic function, and calculate the mean absolute error (MAE) of each fit.
We find that the difference between the normalized MAE values from the considered functions is less than 0.7\%. Therefore, the uncertainty connected with the choice of using a linear regression in the operational AEP methodology at monthly time resolution appears minimal. Moreover, through conversations with wind industry professionals, we found that a linear regression based on monthly data is the standard industry approach when performing bankable\footnote{Results are accepted by banks, investors, and so on for use in financing, buying/selling, and acquiring wind plants.} operational AEP analyses.

\subsection{Operational AEP Methodology}

Given the lack of existing guidelines for a standard approach for \textit{operational} AEP calculations, we base our methodology on conversations with four major wind energy consultants who represent most of the operational market share in North America. These conversations overwhelmingly revealed the following characteristics for operational AEP analysis, and we follow the same approach in our analysis:
\begin{enumerate}
    \item Wind speed data (measured or modeled) are density-corrected at their native time resolution, using Equation \ref{den_corr}.
    \item Monthly revenue meter data, monthly average availability and curtailment losses, and monthly average wind speeds from a long-term wind resource product are calculated.
    \item Monthly revenue meter data are normalized to 30-day months (e.g., for January, the revenue meter values are multiplied by 30/31).
    \item Monthly revenue meter data are corrected for monthly availability and curtailment (i.e., monthly gross energy data are calculated).
    \item A linear regression between monthly gross energy production and concurrent density-corrected monthly average wind speeds is performed. 
    \item Long-term density-corrected monthly average wind speed is then calculated for each calendar month (i.e., average January wind speed, average February wind speed, and so forth) with a hindcast approach, using 10--20 years of the available long-term reference monthly wind resource data (reanalysis products, long-term reference measurements, etc.).
    \item Slope and intercept values from the regression relationship are then applied to the long-term density-corrected monthly average wind speed data  with the long-term or so-called windiness correction. A long-term data set of monthly (January, February, etc.) estimated gross energy production is obtained.
    \item The resulting long-term monthly gross energy estimates, which are based on 30-day months, are then denormalized to the actual number of days in each calendar month (e.g., for January, the obtained value is multiplied by 31/30).
    \item Long-term estimates of availability and curtailment losses are finally applied to the denormalized long-term monthly gross energy data, leading to a long-term calculation of operational AEP.
\end{enumerate}

In the EIA-923 database, availability and curtailment data are not available. Therefore, in our analysis we omit steps 4 and 9 of the list, and only perform calculations on net energy data.

\begin{figure*}[t]
	\centering
	\includegraphics[width=16cm]{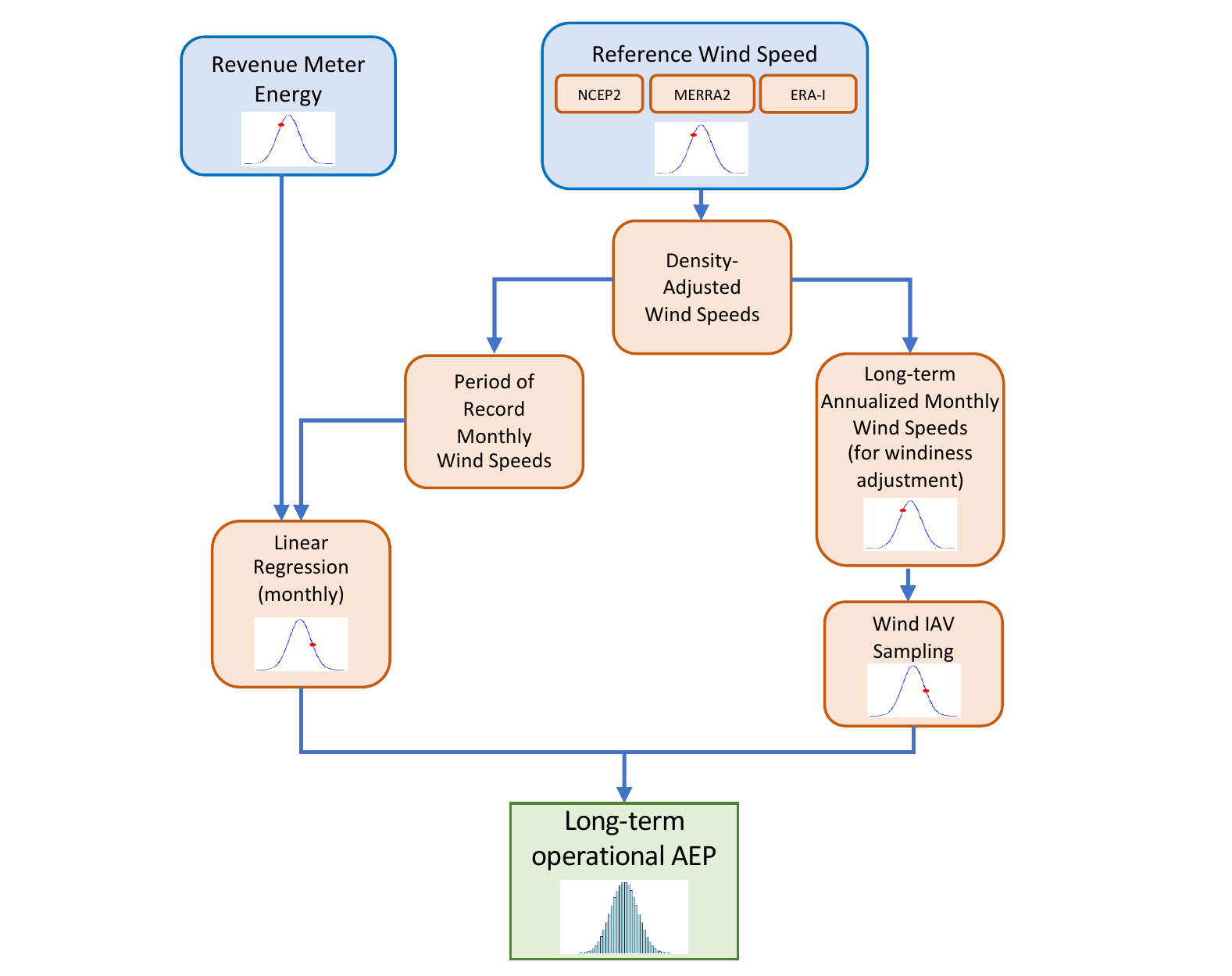}
	\caption{Long-term AEP estimation process using operational data under a Monte Carlo approach; sources of uncertainty and points of Monte Carlo sampling are denoted by probability distribution images. Note: IAV denotes interannual variability.}
	\label{MC_process}
\end{figure*}

\subsection{Monte Carlo Analysis}

To quantify the impact of the single uncertainty components on the long-term operational AEP estimate obtained using the methodology described in the previous section, we implement a Monte Carlo approach. In general, a Monte Carlo method involves the randomized sampling of inputs to, or calculations within, a method which, when repeated many times, results in a distribution of possible outcomes from which uncertainty can be deduced. This is usually calculated as the standard deviation or the coefficient of variation (i.e. standard deviation normalized by mean) of the resulting distribution \citep{iso1995guide,dimitrov2018wind}. Monte Carlo methods have been used in different applications for uncertainty quantification within the wind energy industry, ranging from the prediction of extreme wind speed events \citep{ishihara2015prediction}, to offshore fatigue design \citep{muller2018application}, to economic analysis of the benefits of wind energy projects \citep{williams2008estimating}. Here, we apply this approach to derive a distribution of long-term operational AEP values from which its uncertainty can be calculated. Using a Monte Carlo approach provides a direct estimate of AEP uncertainty by sampling the relevant parameters connected to the various uncertainty components. By contrast, traditional approaches to assessing uncertainty are often less direct. For example, wind resource IAV is often calculated and then converted to AEP uncertainty through an "energy-velocity" (EV) ratio estimated from the wind and energy data. A Monte Carlo approach avoids this intermediate ratio and any uncertainty or error associated with it.

In our analysis, we separately consider five operational-based uncertainty components so that only the sampling of one parameter is performed in each Monte Carlo configuration. The following uncertainty components are included in our proposed Monte Carlo methodology for long-term operational AEP:
\begin{itemize}
    \item Revenue meter measurement error. To incorporate this uncertainty component in the Monte Carlo simulation, we sample monthly revenue meter data from a synthesized normal distribution centered on the reported value and $0.5\%$ imposed standard deviation. In fact, a value of 0.5\% is consistent with what is typically assumed in the wind energy community as revenue meter uncertainty \citep{iec60688,ansic12}. 
    
    \item Reference wind speed data modeling error. Quantifying the uncertainty of the long-term wind resource data used in the operational AEP assessment is challenging because it can vary based on the location, long-term wind speed product used, or instrument from which reference observations are taken. To include this uncertainty component in a systematic way across the 472 locations considered in our analysis, we adopt an ensemble uncertainty approach \citep{taylor2009wind, zhang2015comparison} and use as proxy the variability of the wind resource between different reanalysis products. Therefore, at each Monte Carlo iteration at each site, we randomly select wind resource data from one of the three considered reanalysis products.
    
    \item Linear regression model uncertainty. We adopt a novel way, directly enabled by the use of Monte Carlo, to incorporate this uncertainty component in the operational AEP assessment. We sample the regression slope and intercept values from a multivariate normal distribution centered on their best-fit values and covariance matrix equal to one of the best-fit parameters. The diagonal terms in the covariance matrix are given by the square of the slope and intercept standard errors. For a regression model between an independent variable, $x$, and a dependent variable, $y$, the standard error of the regression is defined \citep{iso1995guide} as:
    \begin{equation}
    e_y = \sqrt{\frac{\sum \left(y_i - \hat{y_i} \right)^2}{n-2}},
    \end{equation}
    where $\hat{y_i}$ is the regression-predicted value for $y_i$ and $n$ is the number of data points used in the regression. The standard error of the regression slope is:
    \begin{equation}
    e_a = \frac{e_y}{\sum \left(x_i - \overline{x_i} \right)^2},
    \label{unc_slope}
    \end{equation}
    and the standard error of the intercept is:
    \begin{equation}
    e_b = e_y \: e_a \sqrt{\frac{\sum x_i^2}{n}}.
    \label{unc_intercept}
    \end{equation}
    $e_a^2$ and $e_b^2$ are the diagonal terms in the covariance matrix of the multivariate normal distribution of regression slope and intercept from which Monte Carlo values are drawn.
    Slope and intercept values are strongly negatively correlated, which is captured by their covariance when performing the linear regression. The off-diagonal terms in the covariance matrix of the multivariate normal distribution constrain the random sampling of slope and intercept values to avoid sampling unrealistic combinations.
    \begin{figure*}[t]
    	\centering
    	\includegraphics[width=16cm]{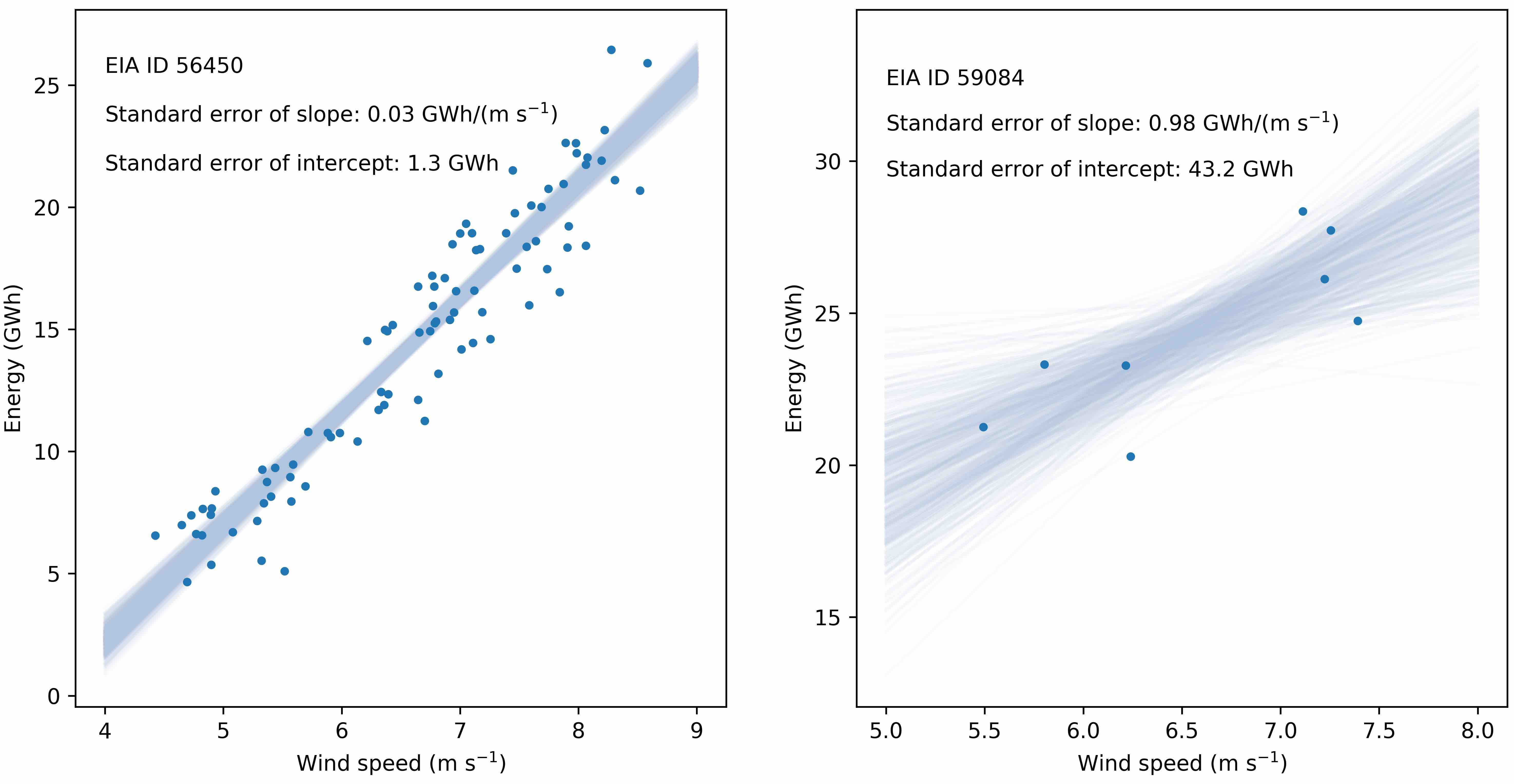}
    	\caption{Sampling set of regression lines corresponding to the slope and intercept values derived from their standard errors in the Monte Carlo approach, for two stations in the EIA data set.}
    	\label{Ex_regression}
    \end{figure*}
    An example of this sampling is shown in Figure \ref{Ex_regression} for two projects of different regression strengths. We sample 500 slope and intercept values from a multivariate normal distribution centered around the best-fit parameters, and with the covariance matrix derived from the standard errors of slope and intercept and their covariance. As shown in Figure 4, the low standard errors found for the leftmost regression relationship constrain the possible slope and intercept values that can be sampled, while the high standard errors in the rightmost regression relationship allow for a much wider sampling.
    
    \item Long-term (windiness) correction uncertainty. We incorporate this component by sampling the number of years (randomly picked between 10 and 20) to use as the long-term wind resource data to which the regression coefficients are applied to derive long-term energy production data (the so-called windiness correction).
    
    \item Wind resource interannual variability (IAV) uncertainty. We incorporate this uncertainty component in the Monte Carlo method by sampling the long-term (reanalysis) average calendar monthly wind speeds (i.e., average January, average February) used to calculate long-term monthly energy production data. The sampling distribution is normal, centered on the calculated long-term average calendar monthly wind speed, and with a standard deviation equal to the 20-year standard deviation of the long-term average monthly wind speed for each calendar month.
\end{itemize}
Each of the listed sources of uncertainty corresponds to a Monte Carlo sampling and is highlighted by a probability distribution in the flowchart in Figure \ref{MC_process}. Note that uncertainty components related to availability and curtailment losses are not considered in our approach because the EIA 923 database does not include measurements of these losses.

To calculate these uncertainty components at each wind plant, we run the Monte Carlo simulation under five different setups, each of them having only a single sampling performed (i.e., either revenue meter, reference wind speed data, IAV, linear regression, or windiness correction). For each component, we run the Monte Carlo simulation 10,000 times. We quantify the impact of each single uncertainty component on the long-term operational AEP in terms of the coefficient of variation of the distribution of operational AEP resulting from the Monte Carlo simulation run. Convergence of the AEP distribution within 0.5\% of the true mean after the 10,000 Monte Carlo runs was verified for all projects, with 95\% confidence.

The code used to perform the AEP calculations is published and documented in NREL's open-source operational assessment software, OpenOA.\footnote{https://github.com/NREL/OpenOA} Calculations were performed on Eagle, NREL's high-performance computing cluster. Specifically, each wind plant was assigned a different processor and run in parallel. Given the general simplicity of the AEP method used here, computational requirements were moderate despite the 50,000 simulations (10,000 runs x 5 uncertainty setups) required for each wind plant.

\subsection{Combination of Uncertainty Components}

Once the contribution from each uncertainty component to the long-term operational AEP uncertainty has been quantified, the different components need to be combined to obtain the total AEP uncertainty. As stated in the Introduction, it is common practice for wind energy consultants to assume that all uncertainty components are uncorrelated, and combine them using Equation \ref{unc_unc} to obtain $\sigma_{\text{tot,uncorr}}$. To test the validity of this assumption, we apply Equation \ref{unc_unc}, in which each of the five considered uncertainty components $\sigma_i$ is quantified as the coefficient of variation of the corresponding operational AEP distribution obtained by running the Monte Carlo simulation with a single sampling performed. We note that the same values of $\sigma_{\text{tot,uncorr}}$ would be obtained by running the Monte Carlo simulation with, at each iteration, all of the five samplings performed, independently from each other.

We contrast the total AEP uncertainty calculated assuming uncorrelated components with what we obtain by taking into account these correlations in the calculation. Following the guidance in \citet{iso1995guide}, we combine the various uncertainty components and calculate the total long-term operational AEP uncertainty for each wind plant as:
\begin{equation}
    \sigma_{\text{tot,corr}} = \sqrt{\sum_\text{i=1}^\text{N} \sigma_\text{i}^2 + 2 \sum_\text{i=1}^\text{N-1} \sum_\text{j=i+1}^\text{N} R_\text{ij} \sigma_\text{i}  \sigma_\text{j} }
    \label{unc_corr}
\end{equation}
where, in our analysis, $N = 5$ and $R_\text{ij}$ is the correlation coefficient between each pair of uncertainty components calculated from the results obtained for all 472 wind plants considered in the analysis.

The comparison between $\sigma_{\text{tot,uncorr}}$ and $\sigma_{\text{tot,corr}}$ will give insights into the error arising from ignoring the correlations existing between the various uncertainty components.

\section{Results}

\subsection{Operational-Based AEP Uncertainty Contributions}

\begin{figure}[t]
	\centering
	\includegraphics[width=8.3cm]{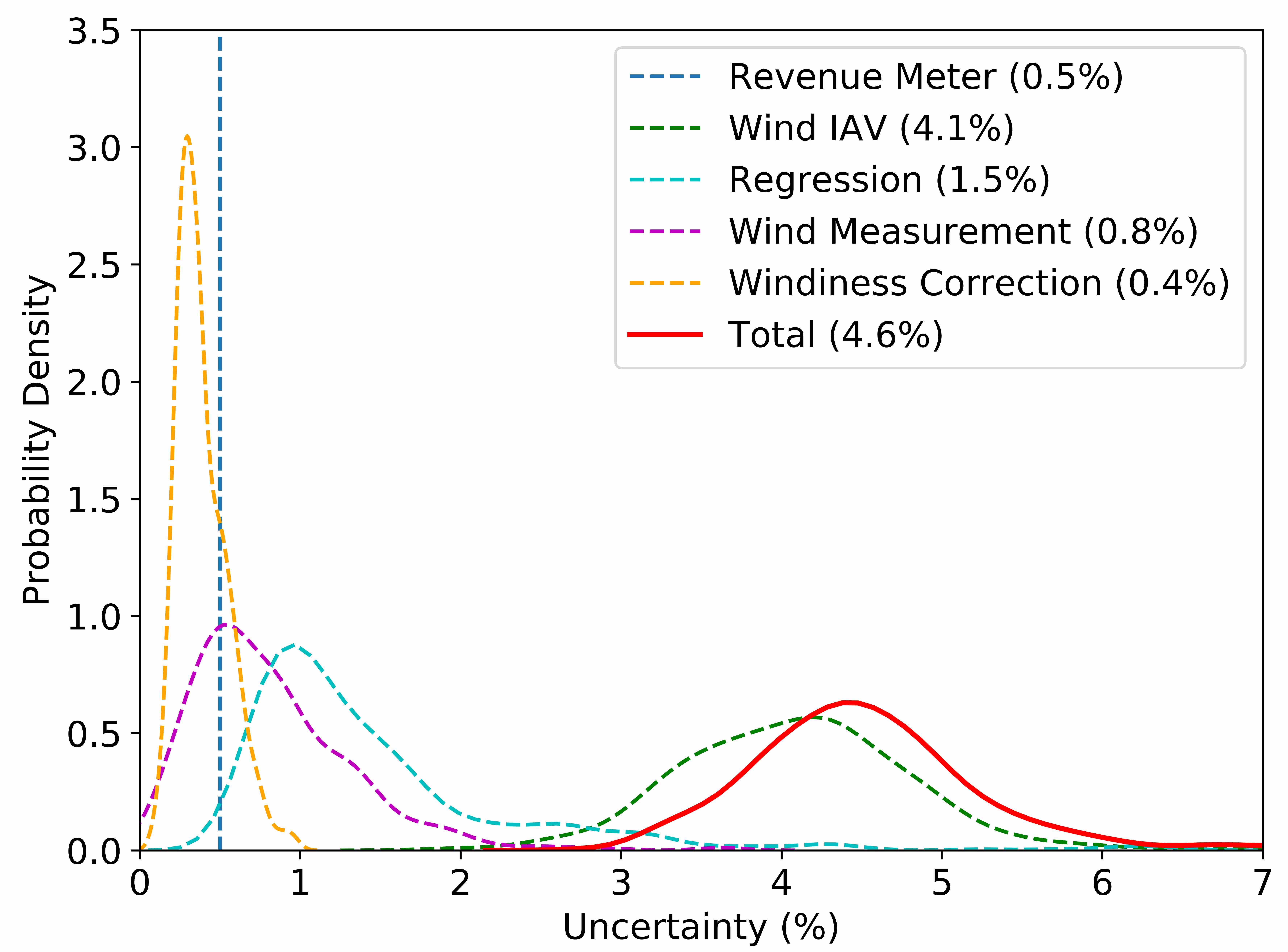}
	\caption{Operational-based AEP uncertainty distributions across projects for the different uncertainty components; mean values across projects are shown in the legend. Uncertainty values are quantified as the percent coefficient of variation of the long-term operational AEP distribution.}
	\label{unc_dist}
\end{figure}

Distributions of each uncertainty component, expressed in terms of the percent coefficient of variation of the resulting AEP distributions, across all 472 wind plants are shown in Figure \ref{unc_dist}
Uncertainty connected to wind resource IAV is found to contribute the most (average 4.1\% across all wind plants). The uncertainty in the linear regression model has the second-largest contribution (1.5\%), followed by the uncertainty of the reference wind speed data (0.8\%; here, of the reanalysis products), and revenue meter data (here, imposed at 0.5\%). The long-term windiness correction has the smallest uncertainty component (0.4\%). Therefore, the number of years used for the long-term windiness correction does not have a large impact on the overall uncertainty in operational AEP, at least for the sampled range of 10--20 years. Using as few as 10 years seems sufficient to give stability to the long-term AEP estimate and adding additional years does not provide a significant reduction in the uncertainty connected with the long-term estimate. As already mentioned in Section 2, these results are obtained for wind plants in mostly simple terrain and with a moderate-to-strong correlation between reanalysis wind resource and wind energy production and, therefore, with an overall low operational AEP uncertainty. We acknowledge that the inclusion of wind plants with a weaker correlation with the reanalysis products would modify the relative contribution of the various uncertainty components (e.g., the importance of the regression uncertainty would increase).

\subsection{Correlation Between Operational-Based AEP Uncertainty Components}

\begin{figure*}[t]
	\centering
	\includegraphics[width=12cm]{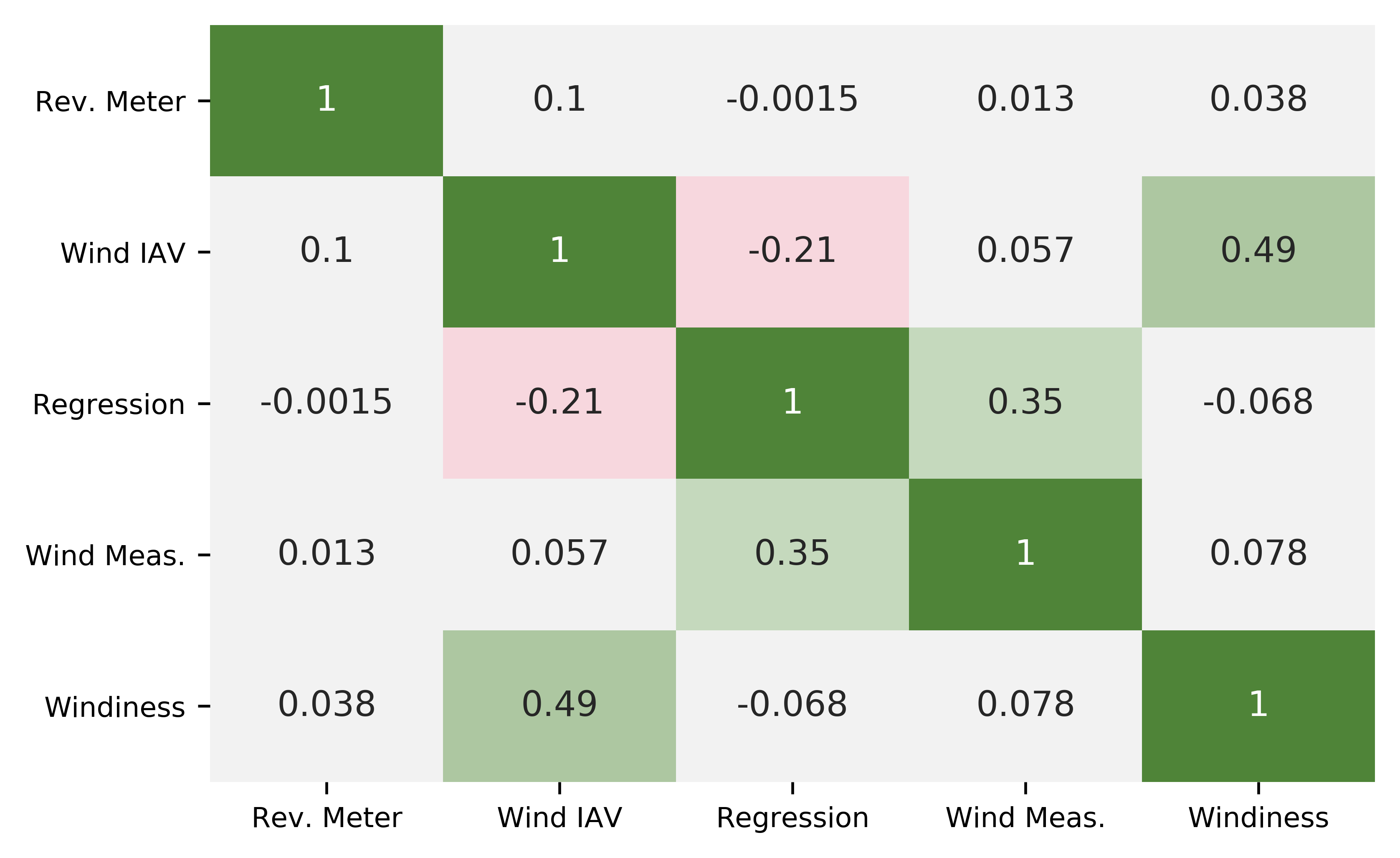}
	\caption{Correlation coefficient heat map between operational AEP uncertainty components, as calculated from the results of the Monte Carlo approach applied at the 472 wind plants considered in the analysis. Note: “Rev.” denotes “Revenue”.}
	\label{unc_corr_heatmap}
\end{figure*}

To be able to assess the validity of the uncorrelated assumption when combining different uncertainty components, we assess potential correlations between uncertainty components by analysing the Pearson's correlation coefficients $R_{ij}$ (needed in Equation \ref{unc_corr} to calculate $\sigma_{\text{tot,corr}}$) from each pair of AEP uncertainty components across the 472 wind plants, and we summarize the results in the correlation matrix in Figure \ref{unc_corr_heatmap}.
To assess which of the obtained correlations have statistical significance, we calculate the $p-$value \citep{westfall1993resampling} associated with the ten correlation coefficients. The test reveals that for three pairs of uncertainty components, the probability of finding the \textit{observed} not-zero correlation coefficients if the \textit{actual} correlation coefficient were, in fact, zero ($p-$value), is less than $10^{-5}$. Therefore, the following three correlations have strong statistical significance:
\begin{itemize}
    \item The wind resource IAV and the long-term windiness correction uncertainties are moderately correlated ($R = 0.49$, $p = 1.9 \cdot 10^{-29}$).
    \item The linear regression and reference wind speed data uncertainties are weakly correlated ($R = 0.35$, $p = 2.5 \cdot 10^{-15}$).
    \item The wind resource IAV and the linear regression uncertainties appear weakly negatively correlated ($R = -0.21$, $p = 2.6 \cdot 10^{-6}$).
\end{itemize}

The first correlation noted earlier (wind resource IAV and long-term windiness correction) is explained simply by the fact that both uncertainty components are driven by wind resource variability. At a site with large wind variability, IAV will be large by definition and so will the uncertainty introduced by different lengths of time series used for the long-term AEP calculation.\\

\begin{figure*}[t]
	\centering
	\includegraphics[width=16cm]{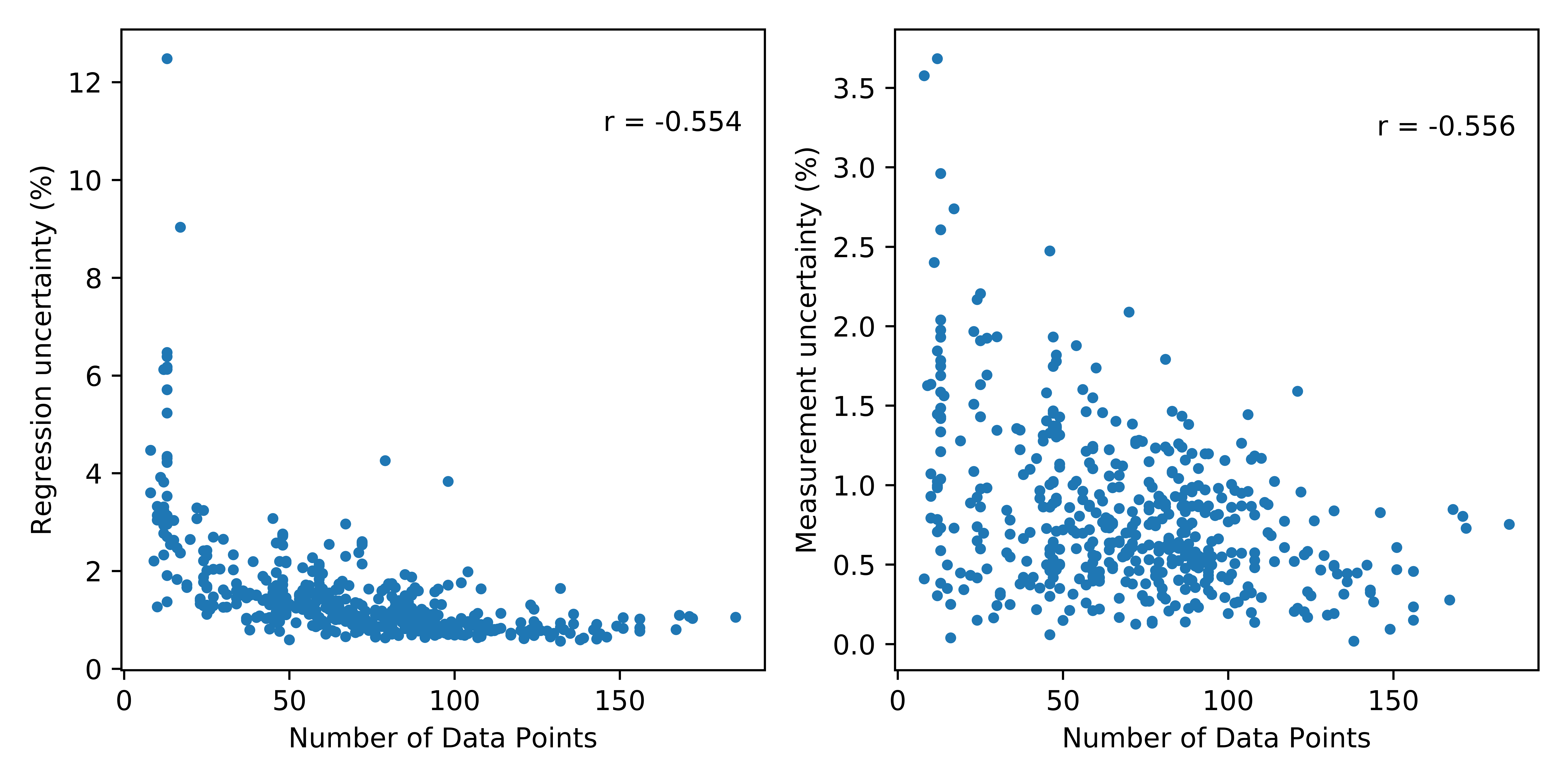}
	\caption{Dependence of linear regression uncertainty and reference wind speed data uncertainty on the number of data points in the period of record, for the 472 projects considered in the analysis.}
	\label{regression_uncertainty}
\end{figure*}

The correlation between linear regression and reference wind speed data uncertainties can be justified given the dependence of both these uncertainty components on the number of data points used in the regression between energy production data and concurrent wind speed data (Figure \ref{regression_uncertainty}).

Both the slope and intercept errors (Equations \ref{unc_slope} and \ref{unc_intercept}), from which the linear regression uncertainty depends (as described in Section 2.3), are inversely proportional to the number of data points so that when a regression is performed on few data points, its uncertainty increases. This dependence is exemplified in Figure \ref{Ex_regression}, where we have compared the sampling sets of regression lines for two stations in the EIA data set: for these two cases, the standard errors of regression slope and intercept for the station with 8 data points (on the right) are 30--50 times larger than what is found for the station with 90 data points (on the left).

\begin{figure*}[t]
	\centering
	\includegraphics[width=16cm]{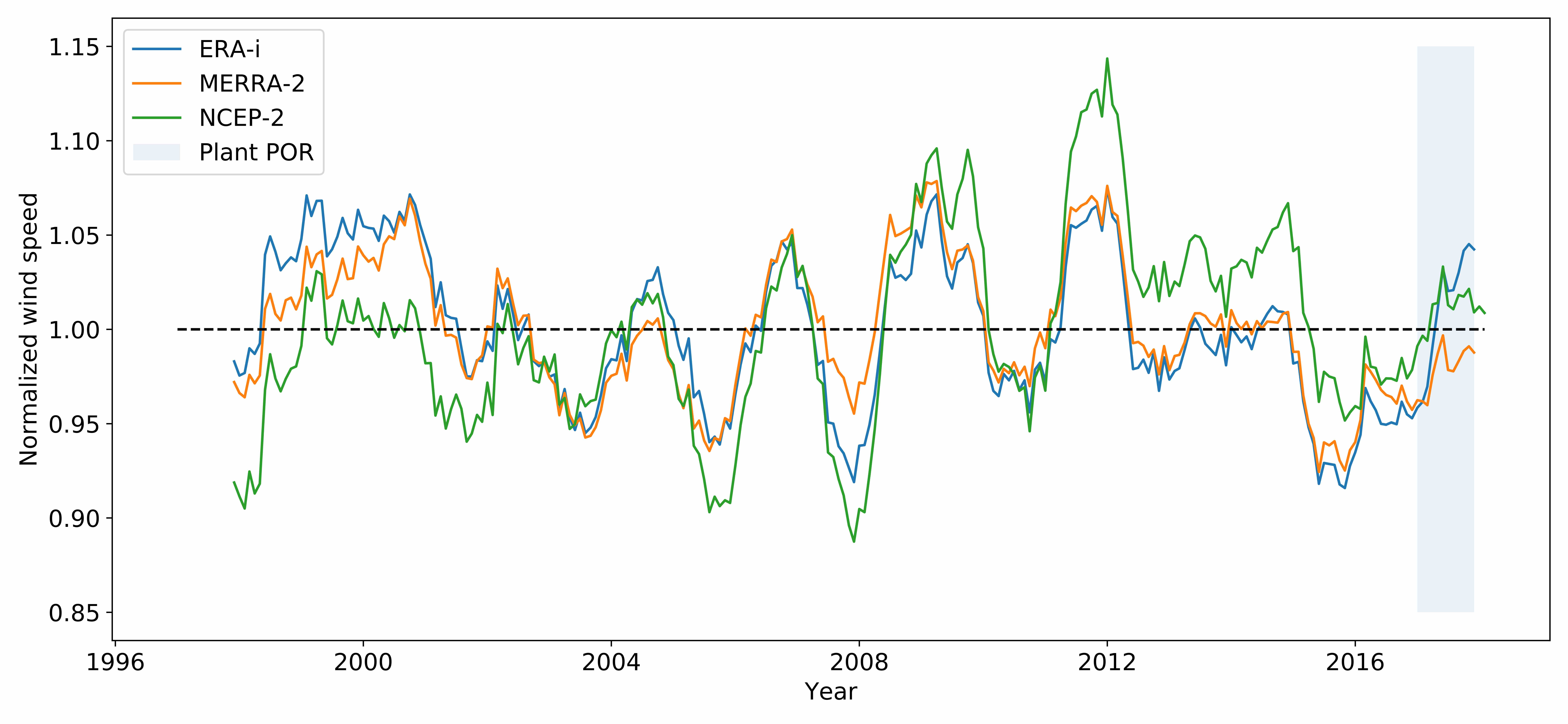}
	\caption{Long-term time series of normalized wind speed for EIA station ID 60502 from the three reanalysis products used in the study. The period of record (POR) for the wind plant is highlighted in light blue.}
	\label{long_ws}
\end{figure*}
\begin{figure*}[t]
	\centering
	\includegraphics[width=16cm]{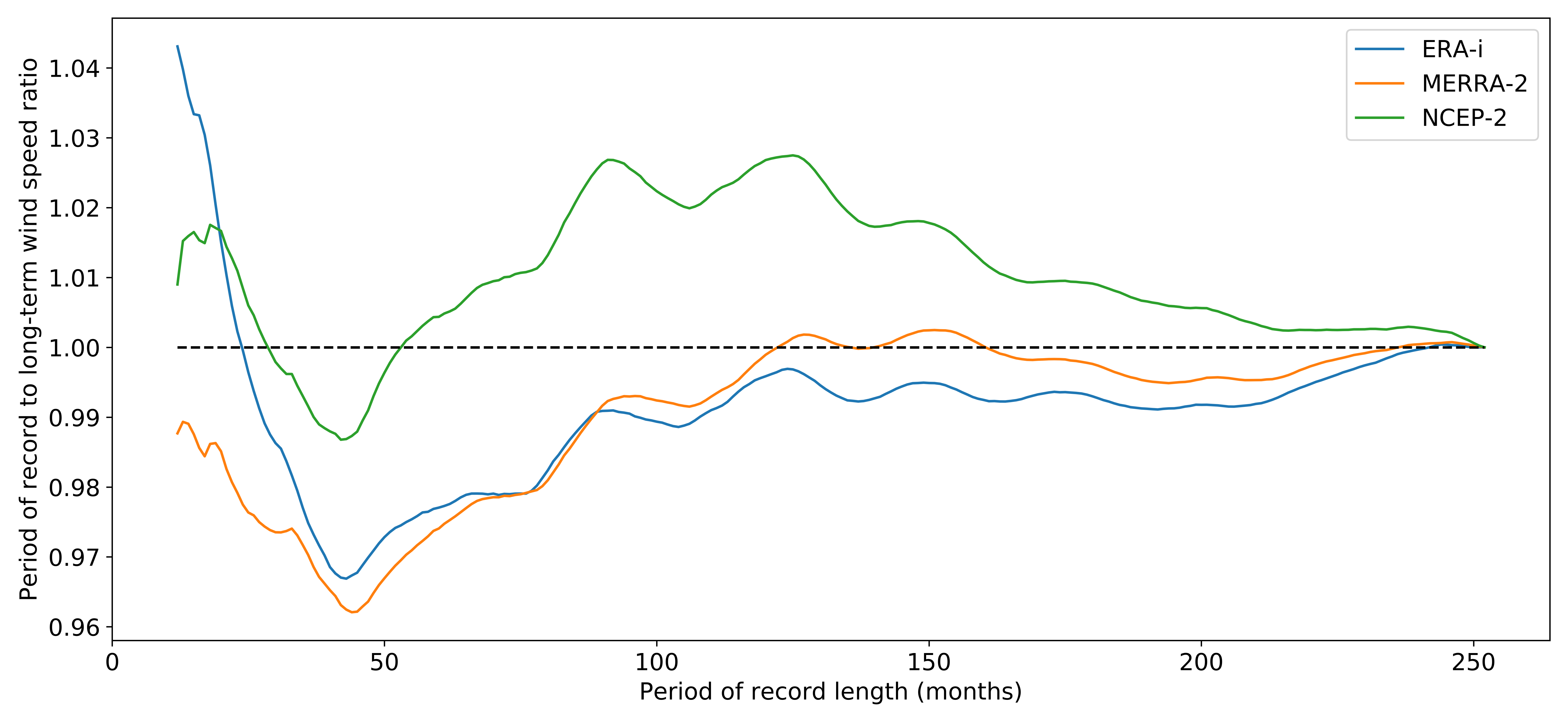}
	\caption{Ratio of wind speed to the long-term, 20-year average for periods of record of different lengths (all ending in December 2017), for EIA station ID 60502, using data from the three reanalysis products in the study.}
	\label{lt_ratio}
\end{figure*}

The number of data points used for the regression also has an impact on the reference wind speed data uncertainty. In fact, short periods of wind plant operation record can lead to different interpretations from the reference wind resource data sets used as to whether that short period of record was above, equal to, or below the long-term average resource. Over a longer period of record, these potential discrepancies between different wind resource data sets (in our case, reanalysis products) tend to average out; therefore, leading to a reduced uncertainty. We illustrate this phenomenon by exploring the long-term trend of the reanalysis products for the wind plant with one of the highest reported reference wind speed data uncertainties (EIA ID 60502 reported 3.7\% reference wind speed data uncertainty). Figure \ref{long_ws} shows the result. The period of record for wind plant operation (shown by a shaded blue area in Figure \ref{long_ws}) was only 12 months. As shown in the figure, the various reanalysis products have very different interpretations of the wind resource in the short period of record relative to the long-term (ERA-I: 4\% above average; MERRA-2: 1\% below average; NCEP-2: 1\% above average). Consequently, the use of each reanalysis product will lead to different magnitudes (both positive and negative) in the long-term windiness corrections, leading to high uncertainty in the resulting operational AEP calculation.
By increasing the period of record (i.e., increasing the number of data points used in the regression), such discrepancies tend to average out. This is illustrated in Figure \ref{lt_ratio}, where we show how the period of record to long-term wind speed ratio varies as we extend the period of record by increasing the number of months while keeping December 2017 as the fixed ending time. For short periods of record, there is considerable deviation of this ratio among the different reanalysis products (i.e., the reference wind speed data uncertainty is high). As the length of the period of record increases, this ratio tends to converge to 1.0, and the spread between the three reanalysis products decreases (i.e., the reference wind speed data uncertainty is low).\\

\begin{figure*}[t]
	\centering
	\includegraphics[width=16cm]{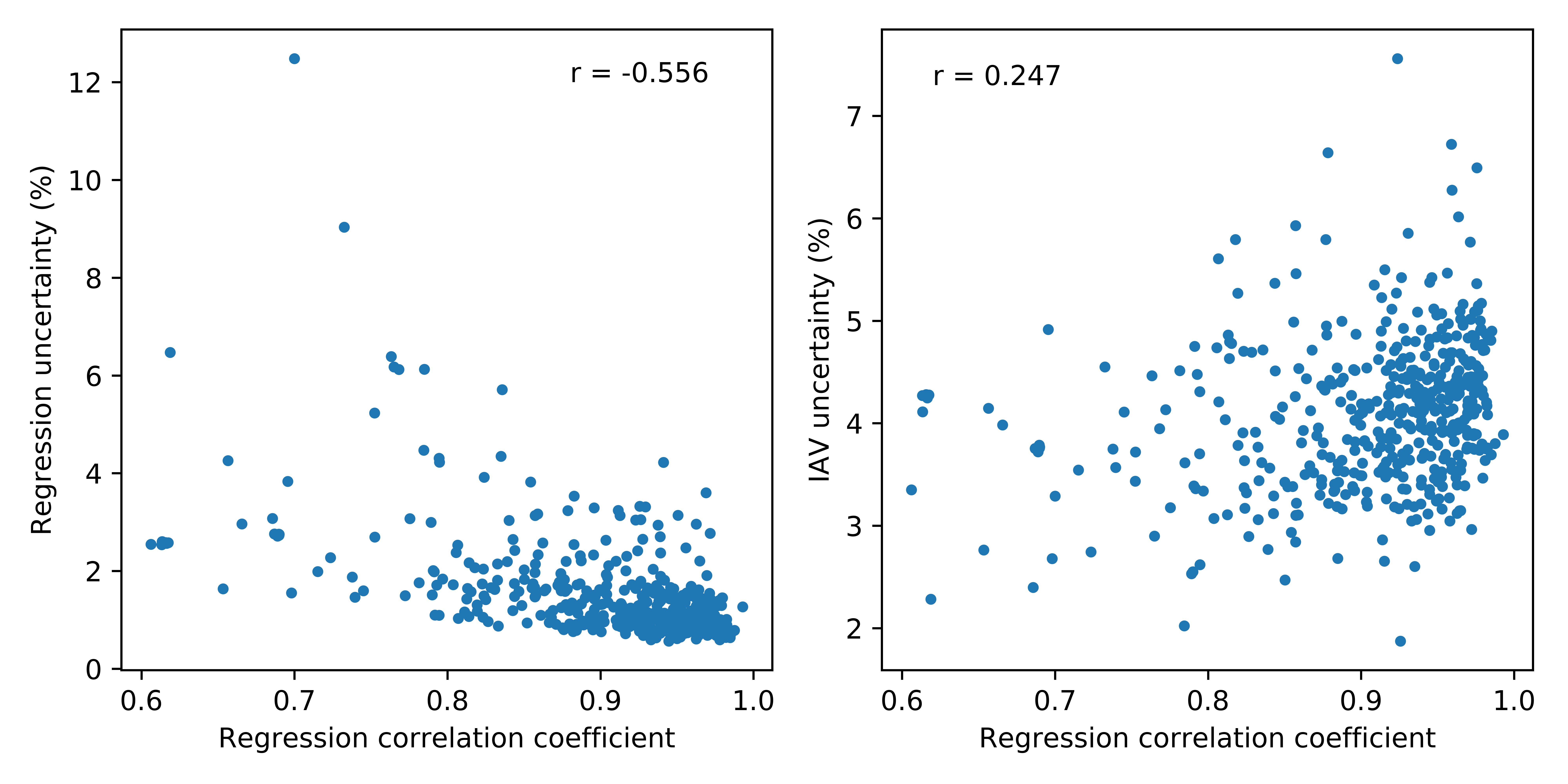}
	\caption{Dependence of linear regression uncertainty and IAV uncertainty on the $R^2$ of the regression between reanalysis wind speed and energy production data.}
	\label{dep_r2}
\end{figure*}

\begin{figure}[t]
	\centering
	\includegraphics[width=8.3cm]{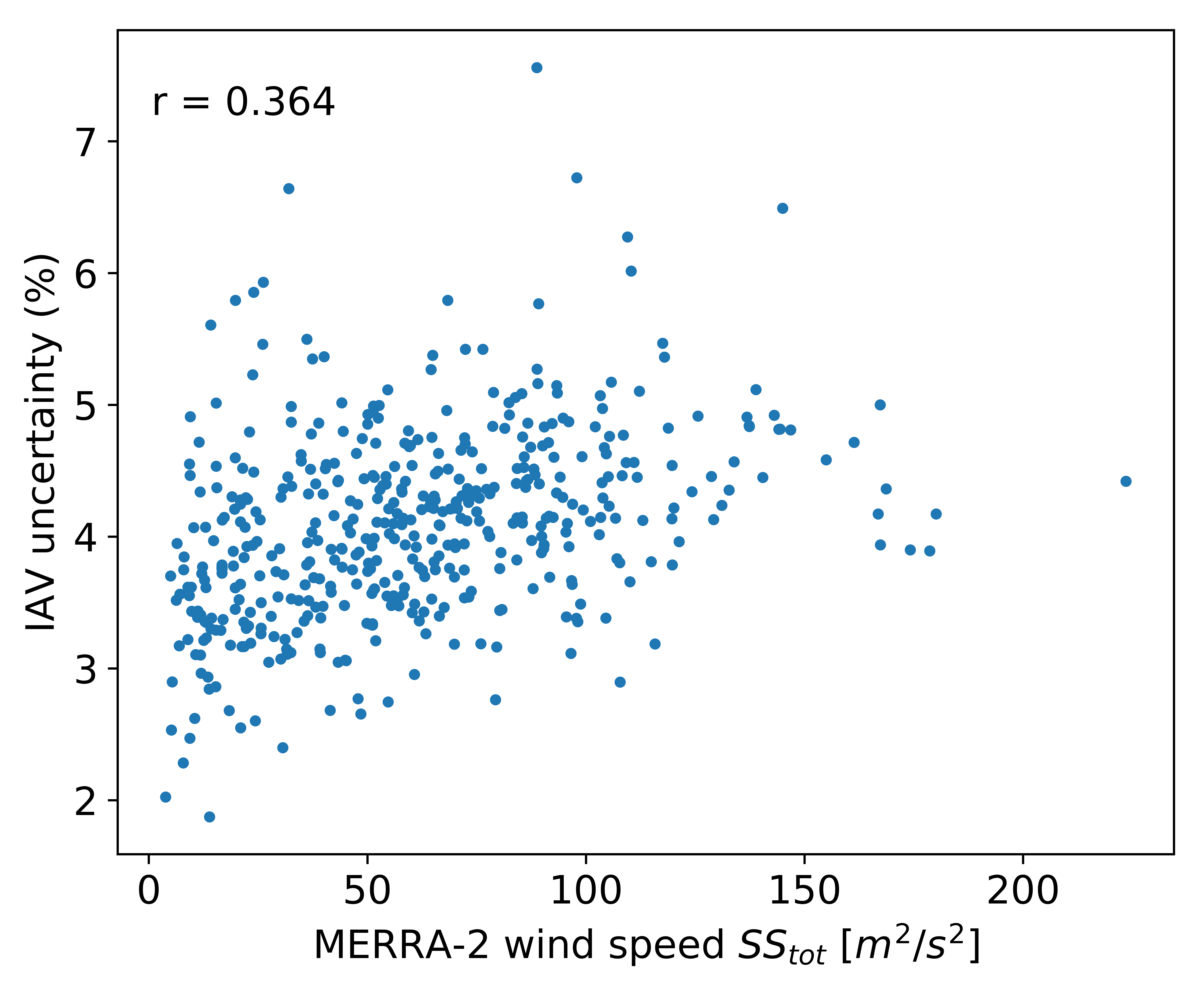}
	\caption{Relationship between IAV uncertainty and the total sum of squares $SS_{\text{tot, WS}}$ of MERRA-2 wind speed data for the 472 projects considered.}
	\label{dep_iav_spread}
\end{figure}

Finally, the (weak) negative correlation between linear regression and wind resource IAV uncertainties is linked to the fact that they respond differently to the $R^2$ coefficient between the reanalysis wind speed and the energy production data (Figure \ref{dep_r2}).
Predictably, the linear regression uncertainty is inversely proportional to the coefficient of determination because a stronger correlation between winds and energy production will lead to a reduced uncertainty of the regression between the two variables.
On the other hand, wind resource IAV uncertainty shows a positive correlation with the regression $R^2$ coefficient. This dependence can be explained because both quantities are positively correlated with the total variance of wind speed or, equivalently, produced energy. Figure \ref{dep_iav_spread} shows the relationship between IAV uncertainty and the total sum of squares $SS_{\text{tot, WS}}$ of reanalysis wind speed (here, using MERRA-2 monthly data), which is proportional to the variance of the data:
\begin{equation}
SS_{\text{tot, WS}} =\sum_i{(WS_i - \overline{WS})^2.}
\end{equation}
A positive correlation between IAV uncertainty and $SS_{\text{tot, WS}}$ emerges.
At the same time, the linear regression $R^2$ coefficient also depends on the variance of the produced energy (and, equivalently, of wind speed) as it is defined as:
\begin{equation}
R^2 = 1 - \frac{SS_{\text{res}}}{SS_{\text{tot}}} 
\label{R2}
\end{equation}
where $SS_{\text{res}}$ is the total sum of the residuals from the linear regression. Equation \ref{R2} shows that when the total sum of squares $SS_{\text{tot}}$ increases, so does $R^2$, thus confirming the positive correlation between $R^2$ and the variance in the data.

\subsection{Comparison Between Total Operational-Based AEP Uncertainty Under Different Assumptions}

After having revealed the correlations existing between different AEP uncertainty components and explained their sources, we can compare the total operational AEP uncertainty calculated when allowing for these correlations (Equation \ref{unc_corr}) with the total uncertainty calculated with the uncorrelated assumption using the conventional sum of squares approach (Equation \ref{unc_unc}). 
\begin{figure*}[t]
	\centering
	\includegraphics[width=16cm]{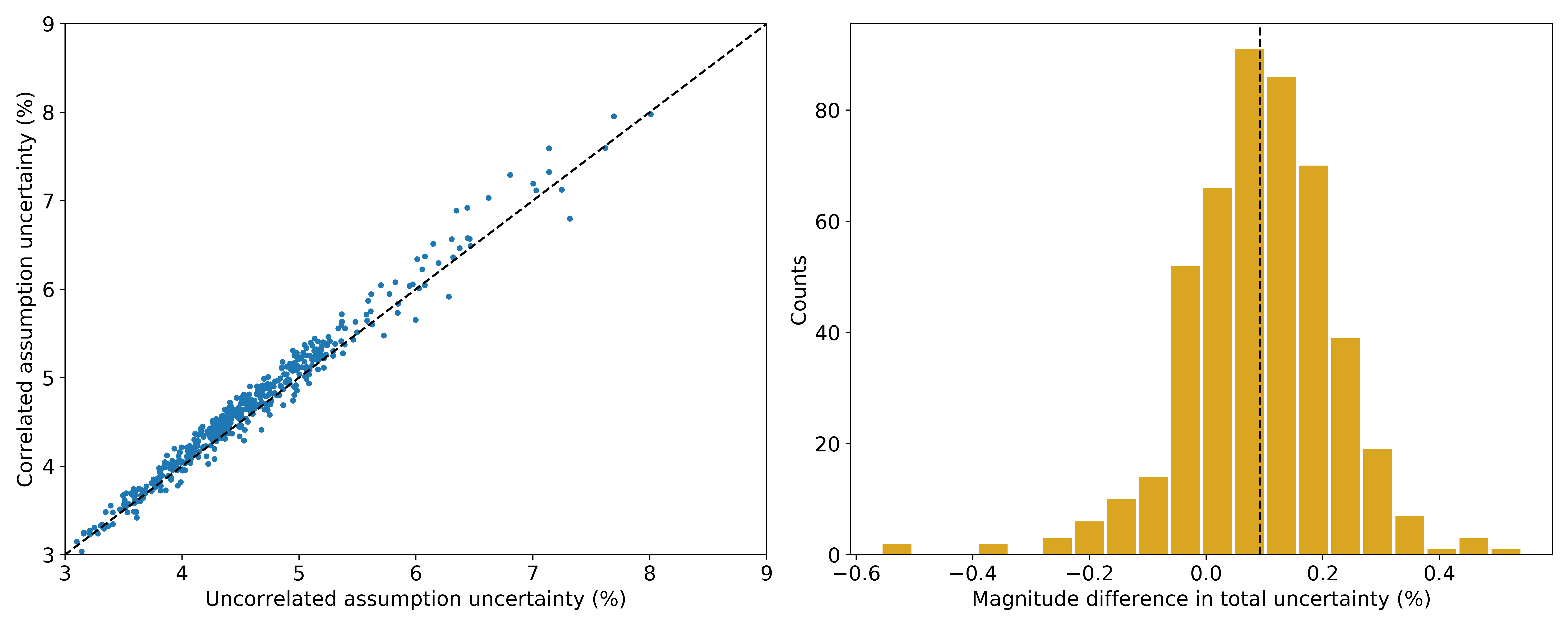}
	\caption{(left) Scatterplot of total operational AEP uncertainty values calculated with and without assuming uncorrelated uncertainty components for the 472 wind plants considered. Uncertainty is quantified as the percent coefficient of variation of the resulting long-term AEP distribution. (right) Histogram of difference $\sigma_{\text{tot,corr}} - \sigma_{\text{tot,uncorr}}$ between the total operational AEP uncertainty calculated considering and ignoring the correlation between its uncertainty components.}
	\label{unc_histogram}
\end{figure*}
Figure \ref{unc_histogram} shows the results of this comparison for the 472 wind plants considered as a scatterplot and also as a histogram of the difference $\sigma_{\text{tot,corr}} - \sigma_{\text{tot,uncorr}}$. A weak bias can be observed with a mean value of $+0.1\%$ in uncertainty difference (and differences up to 0.5\% for specific wind plants). In other words, if correlations between the different uncertainty components are ignored in the calculation method, the whole operational AEP uncertainty is then, on average, slightly underestimated.

This difference can be explained by comparing the contributions $R_\text{ij} \overline{\sigma_\text{i}}  \overline{\sigma_\text{j}}$ from the various uncertainty pairs in Equation \ref{unc_corr} averaged over the 472 considered wind plants.
\begin{figure*}[t]
	\centering
	\includegraphics[width=16cm]{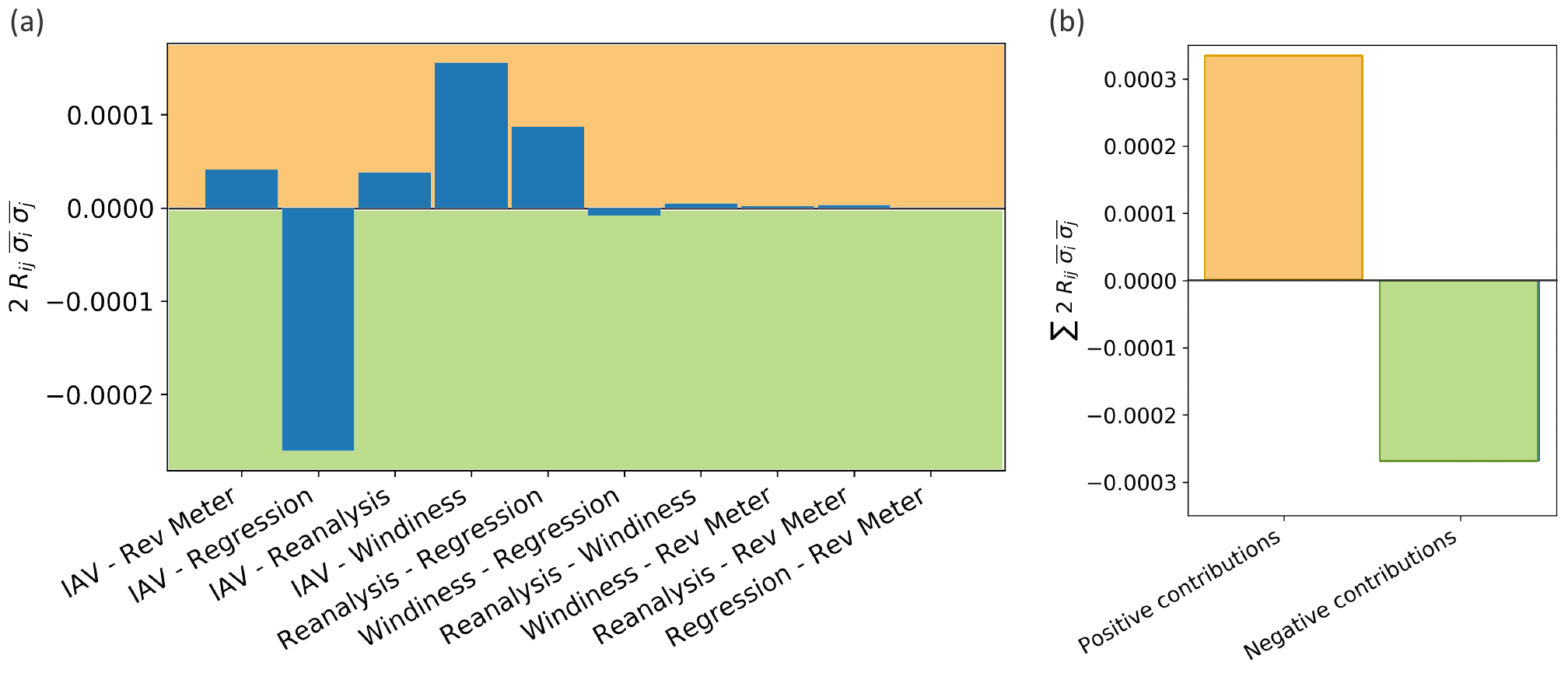}
	\caption{(a) Average (across 472 wind plants) contribution of the correlation between single uncertainty pairs to the total operational AEP uncertainty, according to Equation \ref{unc_corr}. (b) Comparison of the total contribution from positively and negatively correlated uncertainty pairs, computed by summing the contributions shown in panel (a).}
	\label{Pos_neg_corr}
\end{figure*}
Figure \ref{Pos_neg_corr}a shows the mean magnitude (across all wind plants) of these contributions for all of the considered uncertainty pairs. The negative correlation between IAV and linear regression has the largest single impact because this correlation involves the two largest uncertainty components (Figure \ref{unc_dist}). However, the sum of the contributions from all of the positive correlations exceeds the sum of the contribution from the negatively correlated components (Figure \ref{Pos_neg_corr}b), thus resulting in the overall average increase in total operational AEP uncertainty when the correlations are taken into account in the calculation.

\conclusions

Financial operations related to wind plants require accurate calculations of the annual energy production (AEP) and its uncertainty prior to the construction of the plant and, often, in the context of its operational analysis. As wind energy penetration increases globally, the need for techniques to accurately assess AEP uncertainty is a priority for the wind energy industry. Typically, current industry practice assumes that uncertainty components in AEP estimates are uncorrelated. However, we have shown that this assumption is not valid for the five components that comprise an operational-based uncertainty. We used a Monte Carlo approach to assess AEP; this provides quantitative insights into aspects of the AEP calculation that drive its uncertainty. We have applied this approach using operational data from 472 wind plants, mostly in simple terrain, across the United States in the EIA-923 database, in order to study potential correlations between uncertainty components. Three pairs of uncertainty components revealed a statistically significant correlation: wind resource interannual variability (IAV) and long-term windiness correction (positive correlation); wind resource IAV and linear regression (negative); and reference wind speed data and linear regression (positive). Wind resource IAV and long-term windiness correction uncertainties are correlated because they both depend on wind resource variability. Wind resource IAV uncertainty is correlated with linear regression uncertainty because they are both inversely proportional to the number of data points in the period of record. Finally, reference wind speed data uncertainty and linear regression uncertainty show a negative correlation because they respond oppositely to the $R^2$ coefficient between the (reanalysis) wind speed and energy production data.

Our results show that ignoring these correlations between uncertainty components causes an underestimation of the total operational AEP uncertainty of, on average, about 0.1\%, with peak differences of 0.5\% for specific sites. These differences, though not large, would still have a significant impact on increasing wind plant financing rates. Moreover, we expect differences would become even larger for sites characterized by a more complex wind flow. Therefore, our results suggest that correlations between uncertainty components should be taken into account when assessing the total operational AEP uncertainty.

Additional components of uncertainty in an operational AEP were not considered in our study because of limited reporting in the EIA-923 database. These components include reported availability, curtailment uncertainty, and various uncertainties introduced through analyst decision-making (e.g., filtering high-loss months from analysis and regression outlier detection). Future studies could include the impact of these additional sources of uncertainty on the operational AEP assessment. Moreover, our analysis excluded sites, mostly in complex terrain, with a weak correlation between reanalysis wind resource data and wind power production. Future work could explore the magnitude of operational AEP uncertainty and the correlation between its components for such complex flow regimes. Finally, this study focused on correlations between operational AEP uncertainty components. Future work could explore correlations between the numerous preconstruction AEP uncertainty components (e.g., wake loss, wind speed extrapolation, wind flow model).

\codedataavailability{EIA data used in this study are accessible from \url{https://www.eia.gov/electricity/data/eia923/}. Geographical data of the EIA wind plants are available at \url{https://www.eia.gov/maps/layer_info-m.php}. Software used to assess operational AEP is available from \url{https://github.com/NREL/OpenOA}.}

\authorcontribution{NB and MO are equal contributors to this work. MO performed the AEP estimates on the wind plants considered in the study. NB and MO analyzed the processed data. NB wrote the manuscript, with significant contributions by MO.} 

\competinginterests{The authors declare that they have no conflicts of interest.} 


\bibliographystyle{copernicus}
\bibliography{biblio}

\end{document}